\documentclass[cits]{PoS}

\usepackage{amssymb,fontenc,times,mathptmx,graphicx}

\newcommand{\apj}{ApJ}
\newcommand{\apjl}{ApJ}

\newcommand{\mnras}{MNRAS}

\newcommand{\kms}{km s$^{-1}$}
\newcommand{\vk}{$v_{\rm k}$}
\newcommand{\vesc}{$v_{\rm esc}$}
\newcommand{\tmrg}{$t_{\rm mrg}$}

\newcommand{\fgas}{$f_{\rm gas}$}
\newcommand{\tagn}{$t_{\rm AGN}$}

\title{Recoiling Black Holes in Merging Galaxies: Relationship to AGN Lifetimes, Starbursts, and the M$_{\rm BH}$-$\sigma_*$ Relation}

\ShortTitle{Recoiling Black Holes in Merging Galaxies}

\author{\speaker{Laura Blecha}\\
        Harvard-Smithsonian Center for Astrophysics\\
        E-mail: \email{lblecha@cfa.harvard.edu}}

\author{Thomas J. Cox\\
        Carnegie Observatories\\
        E-mail: \email{tcox@obs.carnegiescience.edu}}

\author{Abraham Loeb\\
        Harvard-Smithsonian Center for Astrophysics \\
        E-mail: \email{aloeb@cfa.harvard.edu}}

\author{Lars Hernquist\\
        Harvard-Smithsonian Center for Astrophysics \\
        E-mail: \email{lhernqui@cfa.harvard.edu}}

\abstract{
Gravitational-wave (GW) recoil of merging supermassive black holes (SMBHs) may influence the co-evolution of SMBHs and their host galaxies. We examine this possibility using SPH/N-body simulations of gaseous galaxy mergers in which the merged BH receives a recoil kick. With our suite of over 200 merger simulations, we identify systematic trends in the behavior of recoiling BHs. Our main results are as follows. (1) While BHs kicked at nearly the central escape speed (\vesc) are essentially ``lost" to the galaxy, in gas rich mergers, BHs kicked with up to $\sim 0.7$~\vesc~may be confined to the central few kpc of the galaxy.  (2) The inflow of cold gas during a gas-rich major merger may cause a rapid increase in central escape speed; in such cases recoil trajectories will depend on the timing of the BH merger relative to the change in~\vesc. (3) Recoil events generally reduce the lifetimes of bright active galactic nuclei (AGN) but may actually extend AGN lifetimes at lower luminosities. (4) Recoiling AGN may be observable via kinematic offsets ($v_{\rm BH} > 500$ km s$^{-1}$) or spatial offsets ($R_{\rm BH} > 1$ kpc) for lifetimes of up to $\sim$ 10 - 100 Myr. (5) Rapidly-recoiling BHs may be up to $\sim$ 5 times less massive than their stationary counterparts. These mass deficits lower the normalization of the $M_{\rm BH}-\sigma_*$ relation and contribute to both intrinsic and overall scatter. (6) Finally, the displacement of AGN feedback by a recoil event causes higher central star formation rates in the merger remnant, thereby extending the starburst phase of the merger and creating a denser, more massive stellar cusp.}

\FullConference{25th Texas Symposium on Relativistic Astrophysics - TEXAS 2010\\
		December 06-10, 2010\\
		Heidelberg, Germany}

\begin{document}

\section{Introduction}
It has recently been demonstrated that ``central" supermassive black holes (SMBHs) may in fact spend substantial time in motion and offset from the galactic center, owing to gravitational-wave (GW) recoil. Numerical relativity simulations of binary black hole (BH) mergers have revealed that GW recoil may impart a kick to a merged BH of up to 4000~\kms~\cite{campan07a}.  Although the actual distribution of GW kick velocities is not known, estimates using optimistic assumptions indicate that more than a third of major galaxy mergers could result in kicks greater than 1000~\kms~\cite{campan07a, baker08, lousto10b, vanmet10}. This fraction would be smaller if the progenitor BH spins are not large, or if the spins could be partially aligned prior to merger via gas processes or GR precession \cite{bogdan07, dotti10, kesden10}. 

Recoiling BHs may be observable as AGN that are either spatially or kinematically offset from their host galaxies \cite[e.g.]{madqua04, loeb07, bleloe08, blecha11,guedes11,sijack10}. To date, several offset AGN candidates have been discovered, though none have yet been confirmed \cite{komoss08, shield09b,civano10,batche10,jonker10}. We also examine possible indirect consequences of recoil. For example, recoil events could introduce scatter into the tight correlations between BH mass and galaxy bulges \cite{volont07,bleloe08,blecha11,sijack10}. The sudden displacement of the central AGN from these merging systems also could affect starburst properties or the gas and stellar distributions \cite{sijack10,blecha11}.

We conduct hydrodynamic simulations of galaxy mergers to follow the dynamics and accretion of recoiling BHs in evolving merger remnant potentials. Due to the many free parameters involved, we undertake a large parameter study with dozens of galaxy merger models and a wide range of kick velocities. We are thus able to observe trends in the behavior of recoiling BHs in different environments. Our simulation methods and merger models are outlined in \S~\ref{sec:methods}. We discuss the trajectories of recoiling BHs in \S~\ref{ssec:traj}. In \S~\ref{ssec:agn}, we examine the total and offset lifetimes of recoiling AGN, and in \S~\ref{ssec:coev}, we examine the effects of GW recoil on the co-evolution of SMBHs and their host galaxies. We conclude in \S~\ref{sec:conclude}.

\section{Methods}
\label{sec:methods}

We simulate galaxy mergers using the smoothed particle hydrodynamics (SPH) code {\footnotesize GADGET-3} \cite{spring05a}. The code includes models for radiative cooling, star formation, and supernova feedback \cite{sprher03}, as well as BH accretion, as described in \cite{spring05b}. In order to model recoiling BHs in these simulations, we allow for an arbitrary kick velocity~\vk~to be aded to the remnant BH at the time of the BH merger, \tmrg. In most of our analysis, this velocity is scaled to the central escape speed at the time of the merger, \vesc(\tmrg). We allow the BH to accrete both from ambient gas via the Bondi-Hoyle formula and from an ejected disk of gas with a time-dependent accretion rate. These prescriptions are described in more detail in \cite{blecha11}.

We have tested a total of 62 different galaxy merger models in which we vary the galaxy mass ratio ($q$), the total galaxy mass, the gas fraction (\fgas), and the orbital configuration. For easier reference to these models throughout the text, we assign each a name given by q[{\em value}]fg[{\em value}][{\em orb}], where ``q" and ``fg" denote the galaxy mass ratio and initial gas fraction, and each letter {\em orb} is identified with a specific orbital configuration.  (High- and low-total-mass models are not discussed explicitly here; see~\cite{blecha11}.) For each model, we simulate both a merger with no recoil kick and a merger with~\vk/\vesc~$ = 0.9$.  For a subset of these models, we also simulate intermediate values of~\vk. We refer the reader to~\cite{blecha11} for the full details of our initial conditions.

\begin{figure}
\resizebox{0.5\hsize}{!}{\includegraphics{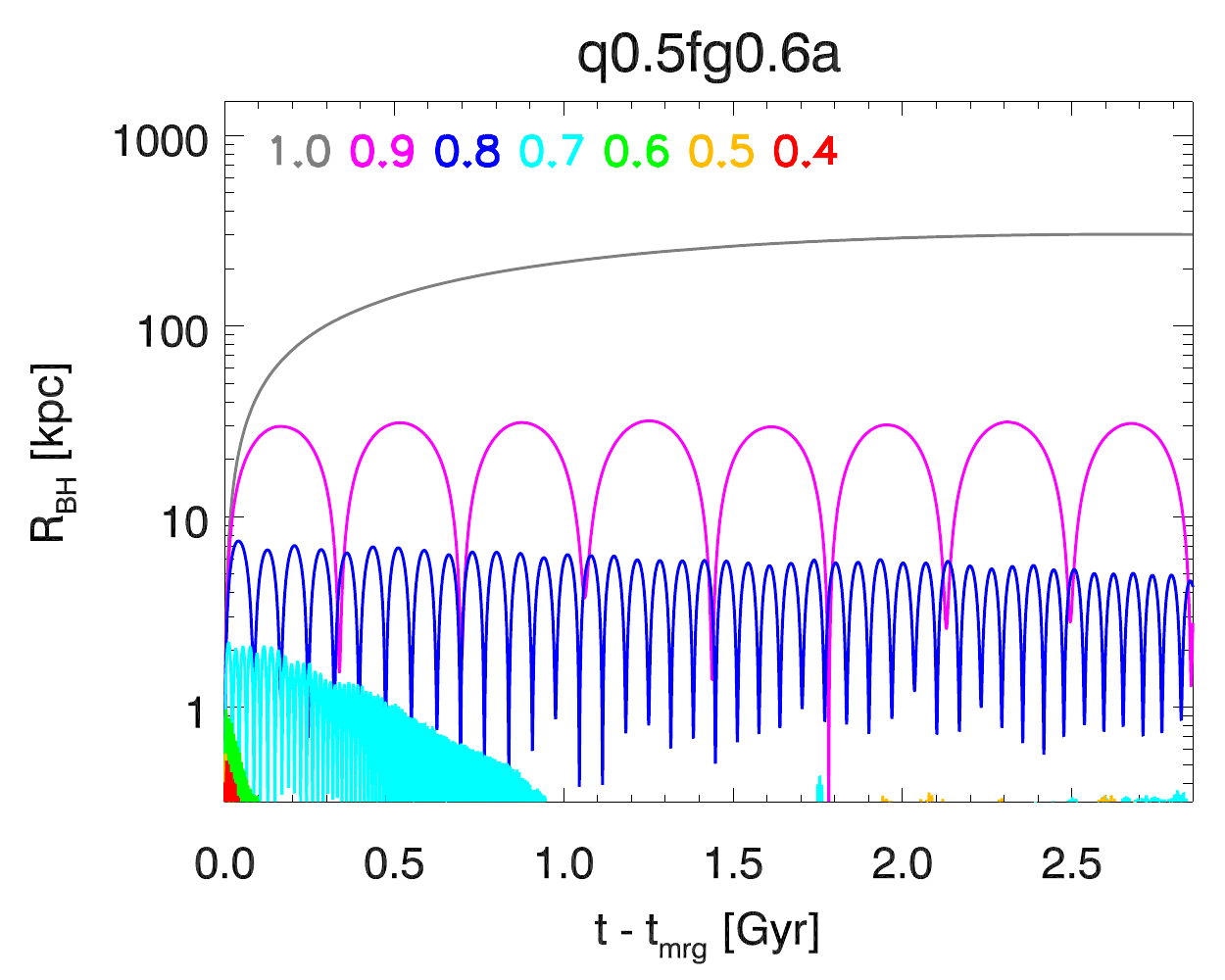}}
\resizebox{0.5\hsize}{!}{\includegraphics{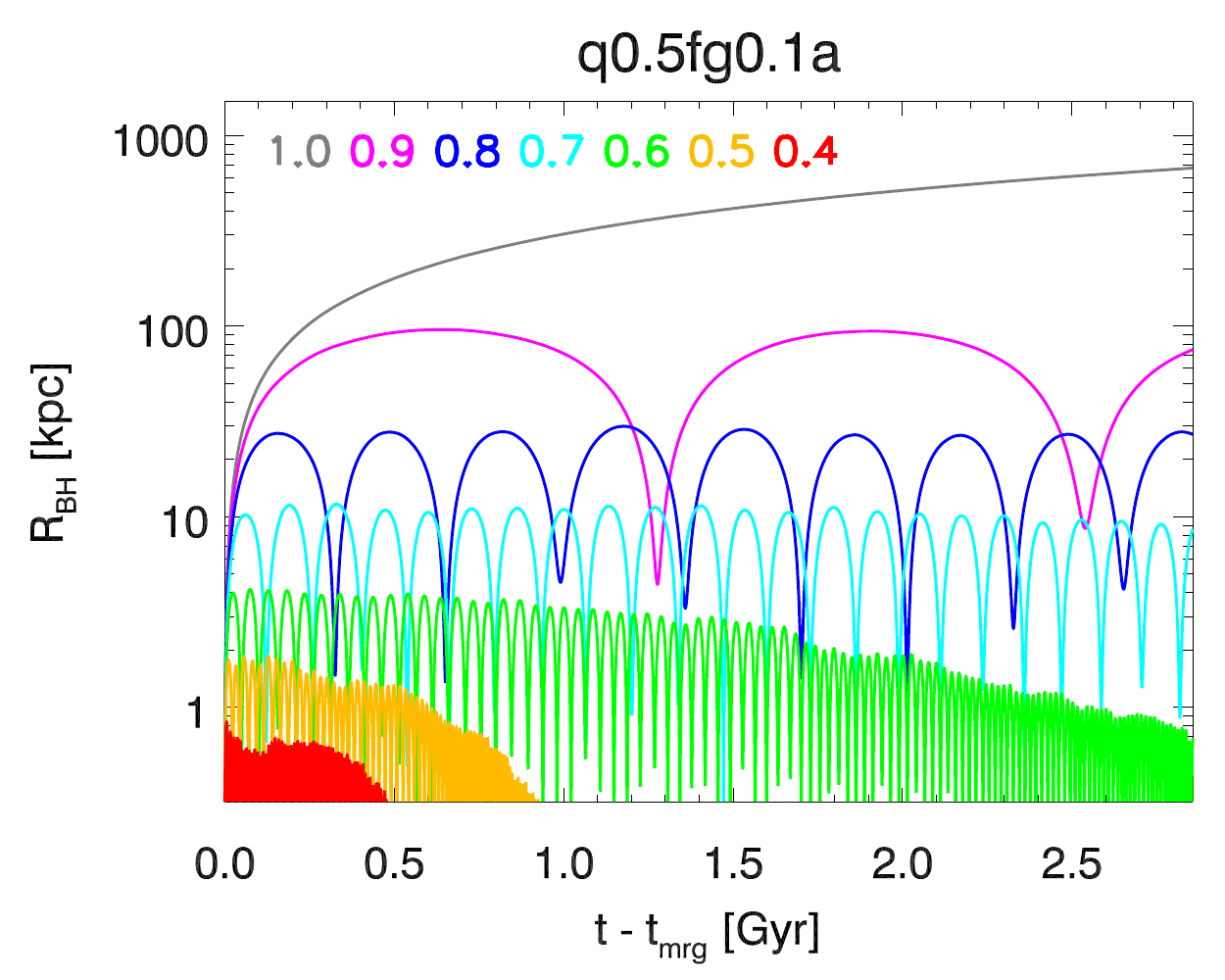}}
\caption[]{Comparison of recoil trajectories in gas-rich (left panel, model q0.5fg0.6a) and gas-poor (right panel, model q0.5fg0.1a) mergers. In each case, the recoiling BH separation from the galactic center is plotted versus time for varying~\vk/\vesc~within a single galaxy model.  Color-coded values of~\vk/\vesc~are indicated on the plots. The $x$-axis is the time after the BH merger, $t - t_{\rm mrg}$.}
\label{fig:2traj}
\end{figure}

\section{Results}
\label{sec:results}

\subsection{Recoiling BH Trajectories}
\label{ssec:traj}

Fig.~\ref{fig:2traj} demonstrates that the gas content of galaxies greatly influences recoiling BH trajectories. In the left panel, trajectories with~\vk~$= 0.4-1.0$~\vesc~are shown for a merger in which the progenitor galaxies had 60\% gas initially, while the right panel shows the same set of~\vk/\vesc~trajectories, but for galaxies with 10\% gas initially. In both cases, BH kicked with $0.8 \lesssim$~\vk~$< 1.0$~\vesc~are still on large orbits by the end of the simulation, 2.9 Gyr after the recoil. However, the BH displacement is smaller in the gas-rich case, and BHs with~\vk~$\lesssim 0.7$~\vesc~($640$~\kms~in this merger model) are confined to the central few kpc of the galaxy. Thus, in gas-rich mergers, even relatively large kicks may fail to eject the BH from the central region.

Recoiling BH dynamics also depend indirectly on gas content via the inflow of cold gas into the center of the merger remnant. We find that nearly-equal-mass, gas-rich mergers may experience a rapid increase in central~\vesc~during the final coalescence of galaxies. Because the BH is also expected to merge during this time, the BH dynamics may depend sensitively on the timing of these two events. However, the effect is much smaller for mergers with smaller $q$ and~\fgas.

\begin{figure}
\resizebox{\hsize}{!}{\includegraphics{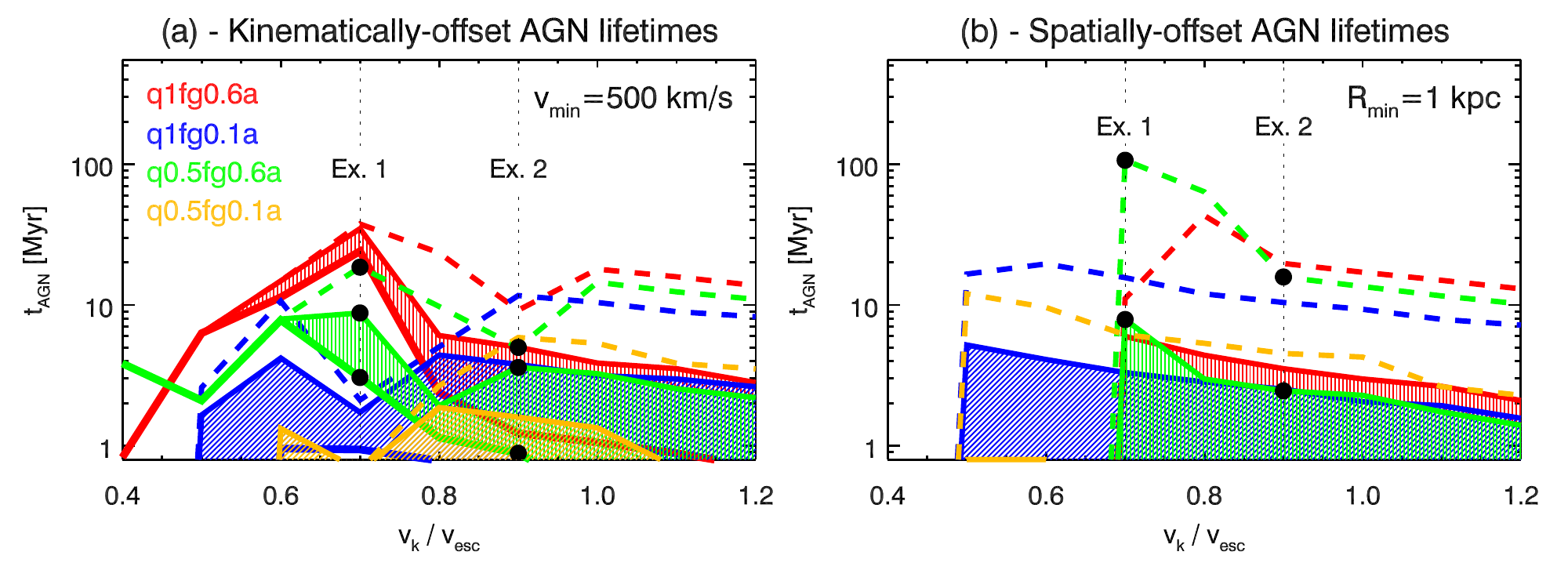}}
\resizebox{0.5\hsize}{!}{\includegraphics{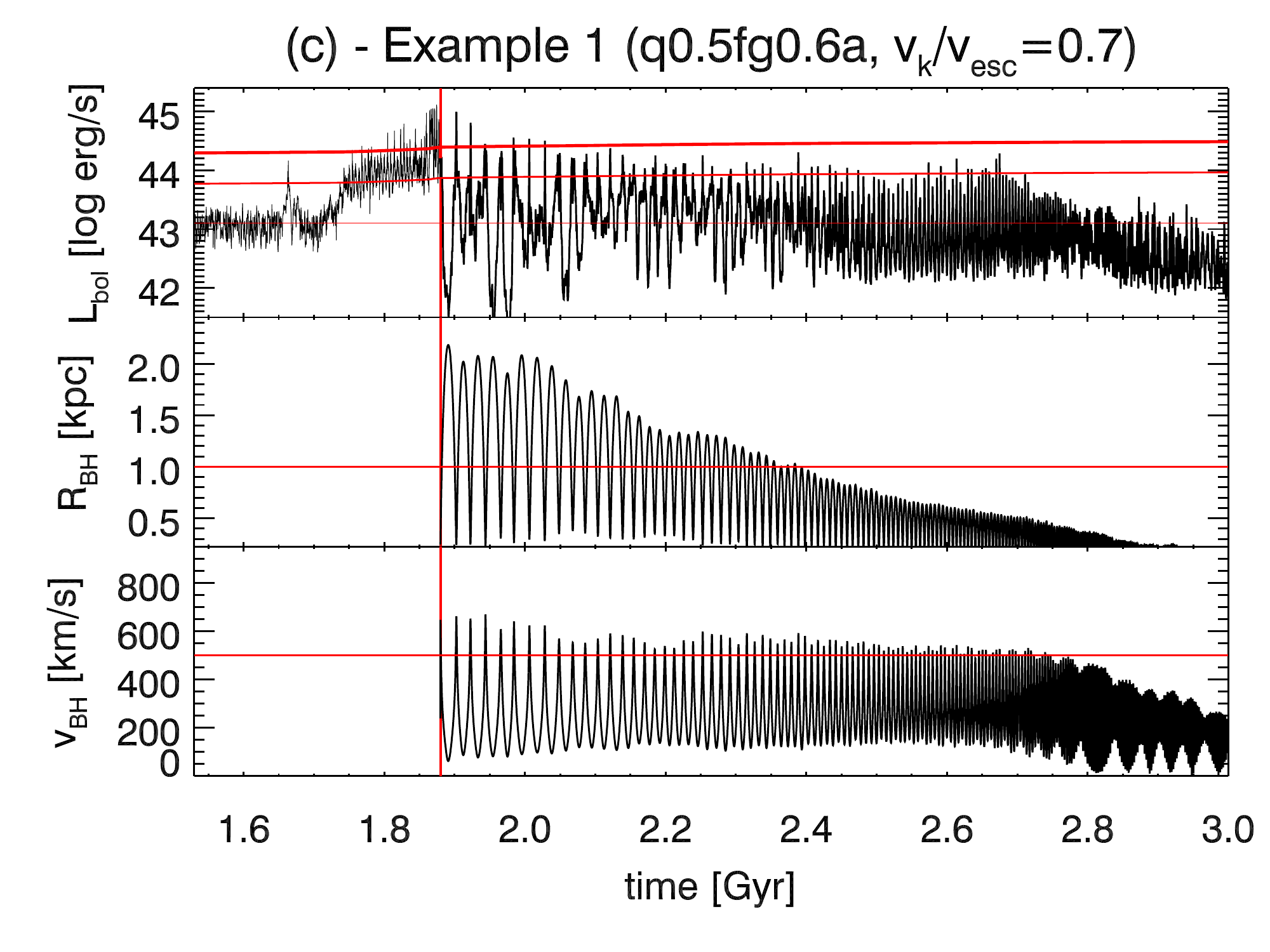}}
\resizebox{0.5\hsize}{!}{\includegraphics{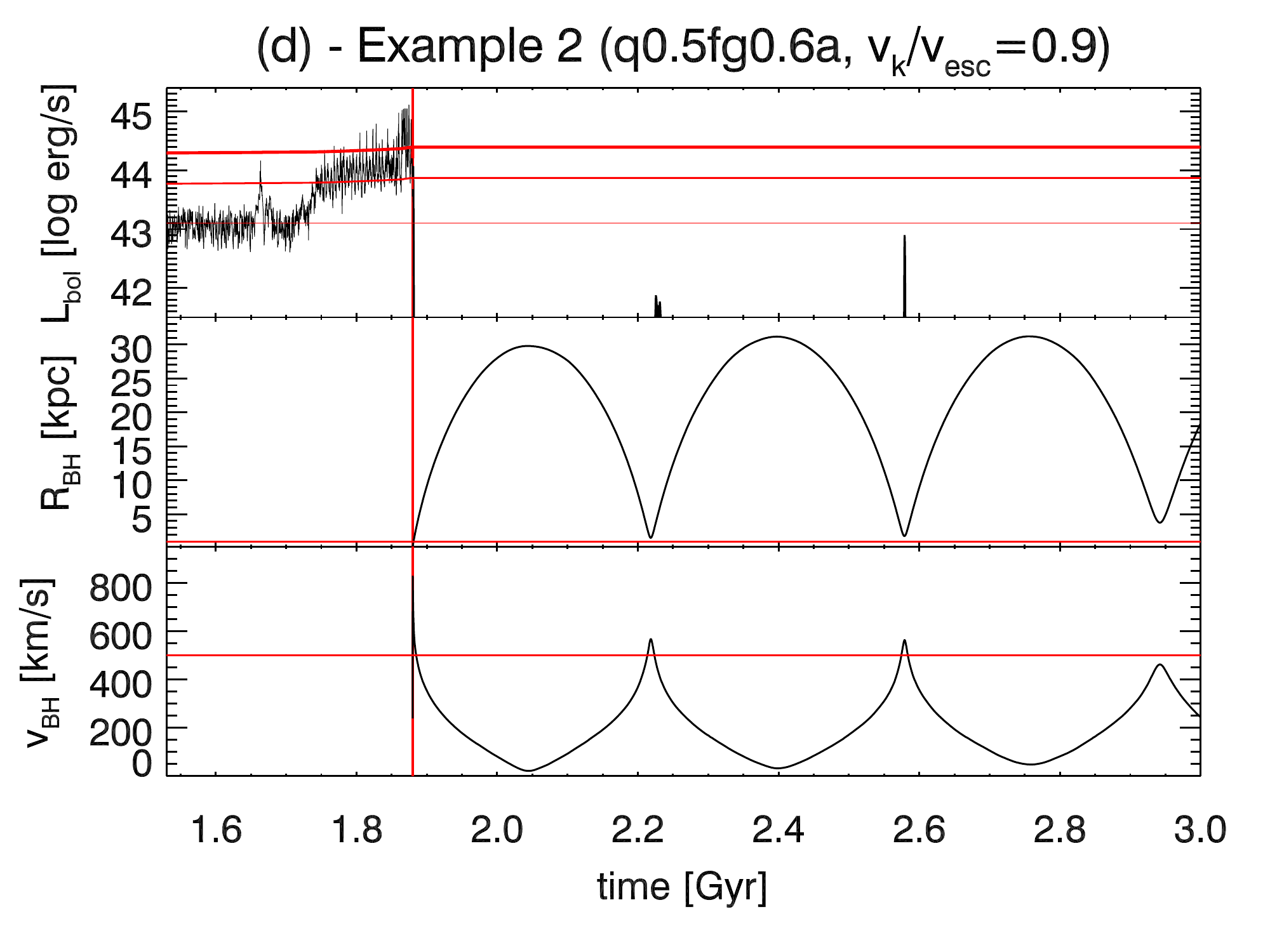}}
\caption[]{Figs.~\ref{fig:tagn_vmin}a \& b show the kinematically- and spatially-offset~\tagn~for $v_{\rm min} = 500$~\kms and $R_{\rm min} = 1$ kpc, respectively. The 4 galaxy models shown are color-coded according to the plot labels. The dashed lines and black circles mark the simulations chosen as examples to illustrate the recoiling AGN luminosity and dynamics in Figs.~\ref{fig:tagn_vmin}c \& d. Fig.~\ref{fig:tagn_vmin}c shows $L_{\rm bol}$, $R_{\rm BH}$, and $v_{\rm BH}$ for the simulation with model q0.5fg0.6a and~\vk/\vesc~$= 0.7$. The vertical red line marks the time of the BH merger. The horizontal red lines in the $L_{\rm bol}$ plot mark the three AGN definitions described in the text. The horizontal red lines in the $R_{\rm BH}$ and $v_{\rm BH}$ plots denote $R_{\rm min}$ \& $v_{\rm min}$, respectively. Fig.~\ref{fig:tagn_vmin}d shows the same data as Fig.~\ref{fig:tagn_vmin}c, but for~\vk/\vesc~$= 0.9$. \label{fig:tagn_vmin}}
\end{figure}

\subsection{AGN Lifetimes}
\label{ssec:agn}

We quantify the observability of recoiling BHs by calculating their lifetimes as either spatially- or kinematically-offset AGN. We use three different definitions of an AGN: $L_{\rm bol} > (10\%\, L_{\rm Edd}$, $3\%\, L_{\rm Edd}$, \& $3.3\times10^9$ L$_{\odot}$), where $L_{\rm bol}$ and $L_{\rm Edd}$ are the bolometric and Eddington luminosities. For kinematically (spatially) offset AGN, we require a BH velocity offset $v_{\rm BH} > v_{\rm min}$ ($R_{\rm BH} > R_{\rm min}$) from the stellar center of mass. Figs.~\ref{fig:tagn_vmin}a \& b show AGN lifetimes (\tagn) for $v_{\rm min} = 500$~\kms~and $R_{\rm min} = 1$ kpc, respectively. 

Figs.~\ref{fig:tagn_vmin}c \& d show the evolution of $L_{\rm bol}$, $R_{\rm BH}$, and $v_{\rm BH}$ for two examples of recoil (\vk/\vesc~$= 0.7$ \& 0.9) in the q0.5fg0.6a merger model. These illustrate two distinct physical scenarios for kinematic offsets. For recoils near~\vesc, the BH carries along a small amount of gas, its luminosity decreasing monotonically with time. For intermediate~\vk/\vesc, the BH may make repeated passages through the central gaseous region. In massive galaxies,~\vesc~may be large enough for BHs on these bound orbits to exceed $v_{\rm min}$ while renewing their fuel supply. Both mechanisms for producing kinematically-offset AGN result in~\tagn~up to a few 10s of Myr in the merger models shown. 

Spatially-offset AGN with $R_{\rm BH} > 1$ kpc (Fig.~\ref{fig:tagn_vmin}b) correspond to an angular separation of $\sim 0.55$ arcsec at $z = 0.1$; these could be resolved with the {\em HST} and {\em JWST} if not observed edge-on. In some cases, spatially-offset AGN (with $R_{\rm BH} > 1$ kpc) may have lifetimes of up to $\sim 100$ Myr, but they are unlikely to appear as bright AGN. Figs.~\ref{fig:tagn_vmin}c \& d illustrate why this is so. In Fig.~\ref{fig:tagn_vmin}c, the largest BH separations correspond to minima in $L_{\rm bol}$. In Fig.~\ref{fig:tagn_vmin}d, the BH displacement is much larger, but the BH has no fuel supply for most of its post-merger evolution.

Finally, we note that in addition to producing offset AGN, GW recoil alters the {\em total} AGN lifetime. Figs.~\ref{fig:tagn_vmin}c \& d show that the bright AGN phase is cut short by the recoil event, but in the \vk/\vesc~$=0.7$ simulation, accretion continues a lower levels for an extended time. In fact, the low-luminosity~\tagn~in this case is actually {\em longer} than for an equivalent stationary BH. This result is generic to many of our intermediate-\vk~simulations.

\begin{figure}
\center{\resizebox{0.45\hsize}{!}{\includegraphics{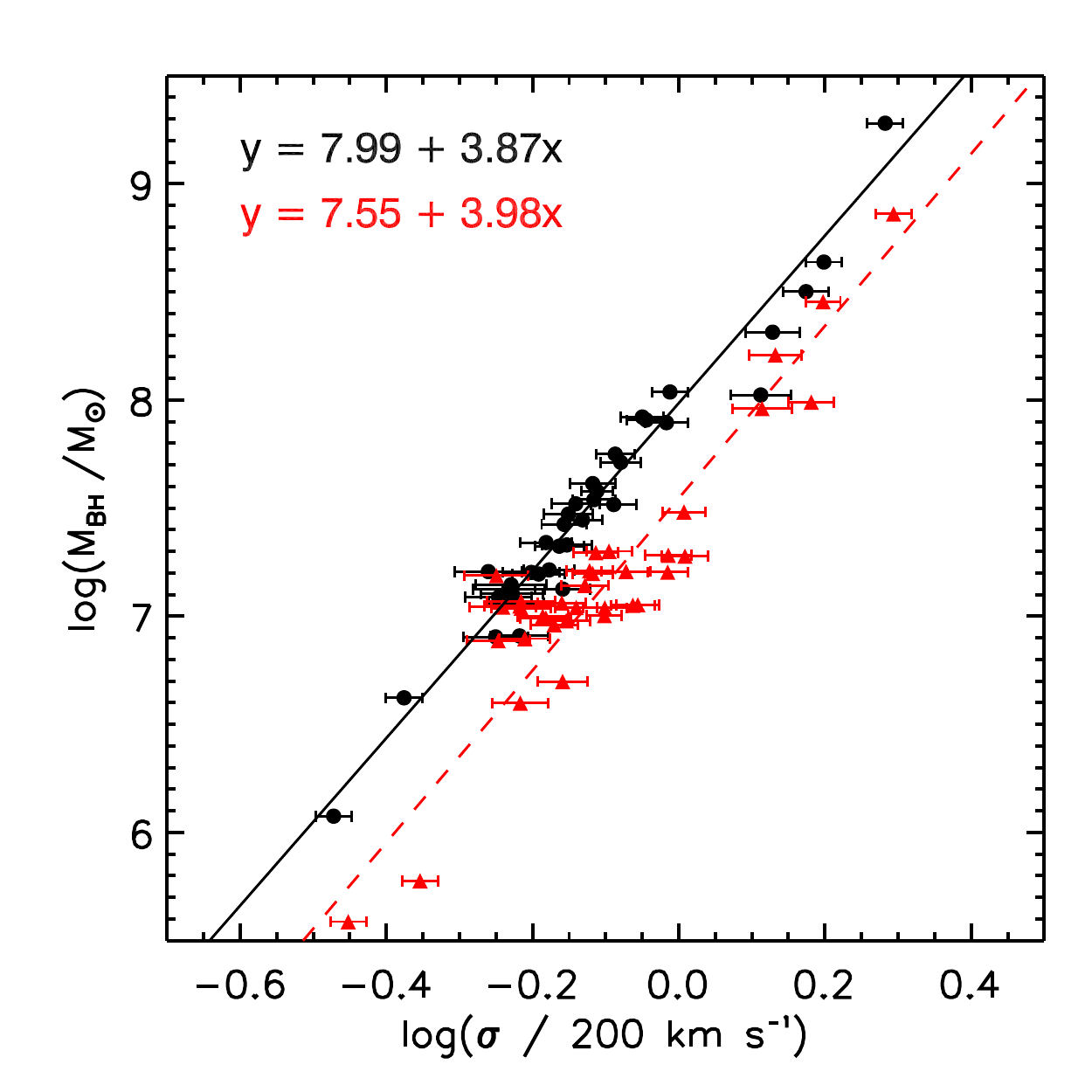}}}
\caption[]{$M_{\rm BH}-\sigma_*$ relation for simulations with no recoil kicks (black circles) and with~\vk/\vesc~$=0.9$ (red triangles).  $M_{\rm BH}$ is the BH mass at the end of the simulation (\tmrg~$+\, 2.9$ Gyr) and $\sigma_*$ is the stellar velocity dispersion averaged over 100 random sight lines, where error bars give the range of sampled values.  The solid black and red dashed lines are least-squares fits to the no-recoil and high-recoil data, respectively. The fit parameters are indicated on the plot. \label{fig:msigma}}
\end{figure}

\begin{figure}
\center{\includegraphics[width=.47\textwidth]{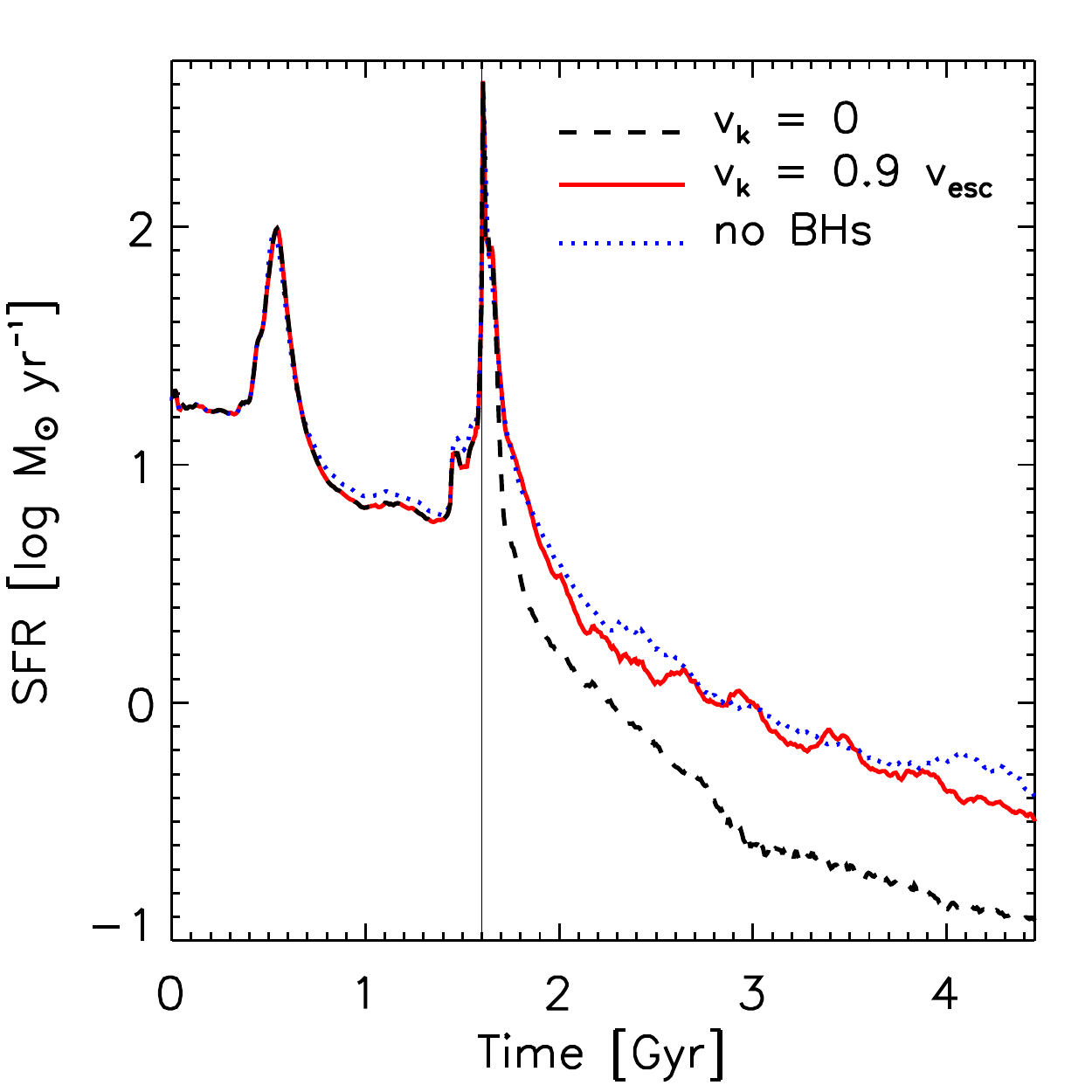}
\includegraphics[width=.47\textwidth]{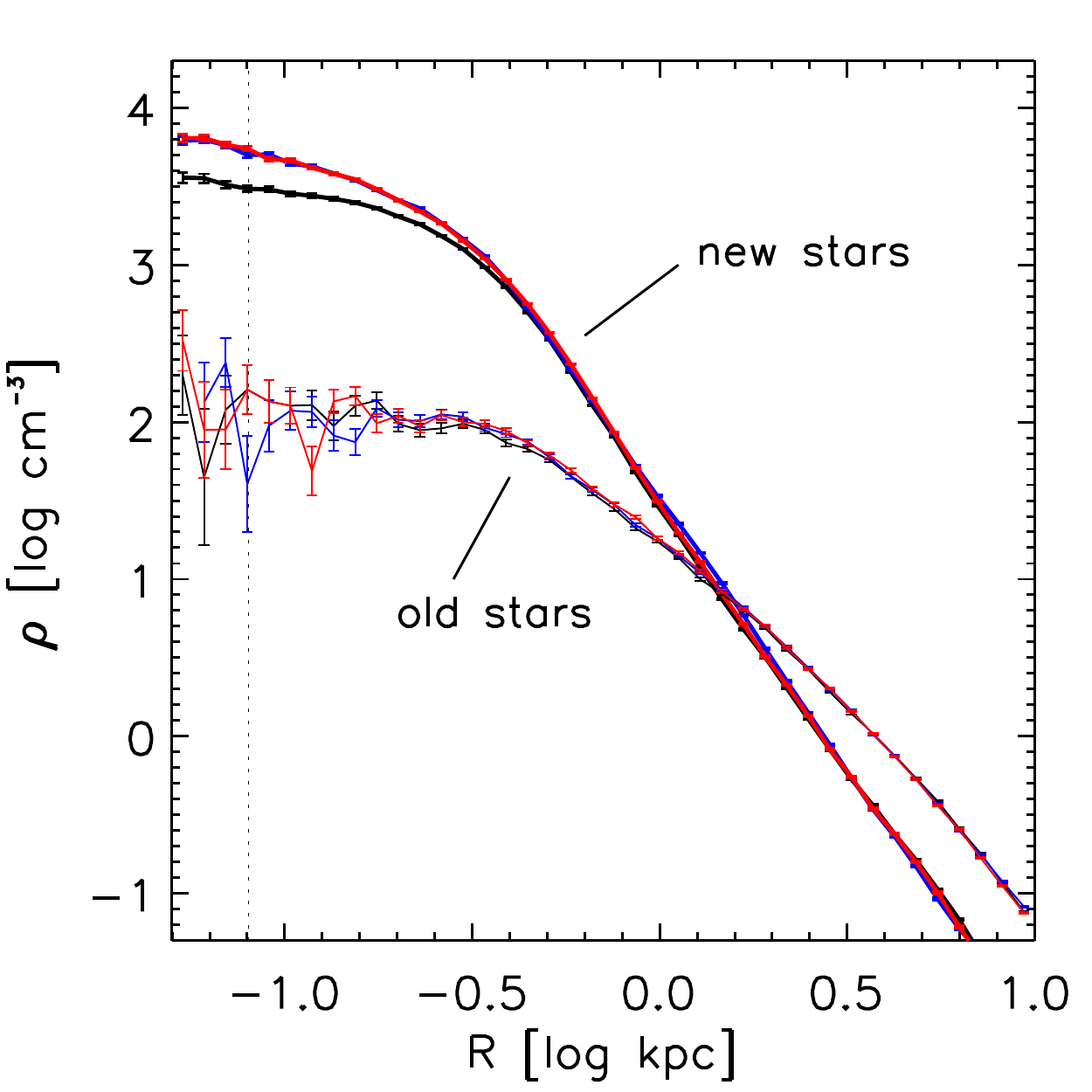}}
\caption[]{Left panel: Total SFR vs. time for the q1fg0.4a merger model.  The simulations shown have~\vk~$= 0$ (black dashed line),~\vk/\vesc~$= 0.9$ (red solid line), or no BHs at all (blue dotted line). The vertical line marks the time of BH merger in the simulations with BHs. Right panel: Final stellar density profiles for the same simulations, with the same color coding.  The thick curves are the ``new stars" that formed during the merger, and the thin curves are the ``old stars" from the progenitor galaxy disks. Poisson error bars are shown for each radius bin.  The vertical dotted line denotes the gravitational softening length (80 pc) in these simulations. \label{fig:comparesfr}}
\end{figure}

\subsection{Co-evolution of BHs and Galaxies}
\label{ssec:coev}

Because recoiling BHs have shorter lifetimes as bright AGN, they generically have smaller final masses than stationary BHs, in some cases by up to a factor of $\sim 5$. In Fig.~\ref{fig:msigma}, we compare the $M_{\rm BH} - \sigma_*$ relation resulting from our sets of simulations with no recoil and with large recoil (\vk/\vesc~$= 0.9$). We stress that the data shown in Fig.~\ref{fig:msigma} are not expected to reproduce the observed $M_{\rm BH} - \sigma_*$ relation~\cite[e.g.]{tremai02}, as we have varied parameters systematically. In particular, galaxies with smaller~\vesc~(and smaller $\sigma_*$) will more frequently experience large~\vk/\vesc~recoils, which could steepen the $M_{\rm BH} - \sigma_*$ slope relative to that shown here. However, by comparing the relative differences between the correlations in Fig.~\ref{fig:msigma}, we gain insight into how individual GW recoil events may contribute to BH/galaxy co-evolution.

Fig.~\ref{fig:msigma} shows that GW recoil lowers the offset of the $M_{\rm BH} - \sigma_*$ relation and increases scatter. While the former effect is somewhat model-dependent, an increase in intrinsic scatter is a necessary consequence of recoil. This is because GW recoil introduces another degree of complexity into the determination of the final BH mass. The scatter in the~\vk/\vesc~$= 0.9$ population in Fig.~\ref{fig:msigma} is almost a factor of two larger than in the~\vk~$= 0$ population (0.24 and 0.13 dex, respectively). In a cosmological framework, scatter would increase even further as some recoiling BHs were replaced by new BHs via subsequent mergers.

In addition, GW recoil can directly affect the central structure of galaxy merger remnants. The left panel of Fig.~\ref{fig:comparesfr} compares the star formation rates (SFRs) of simulations with~\vk/\vesc~$= 0$ \& 0.9, and one with no BHs. The simulation with large recoil clearly has a higher post-merger SFR than the no-recoil simulation, similar to the case with no BHs. This is a direct result of the displacement of AGN feedback energy via GW recoil, which leaves a supply of cold gas in the central region. Owing to this enhanced SFR, a steeper density cusp forms in the merger remnant.  The right panel of Fig.~\ref{fig:comparesfr} shows the final stellar density profiles in these remnants; the newly-formed stars dominate within the central kpc. The simulations with recoil and without BHs have about twice the central density of the stationary-BH case, which corresponds to a 3\% increase in total stellar mass.

\section{Conclusions}
\label{sec:conclude}

We have conducted the first detailed parameter study of GW recoil in hydrodynamic simulations of galaxy mergers. Our large suite of simulations allows us to analyze trends in GW recoil events with varying galaxy merger parameters and recoil velocities. We estimate the observable lifetimes of recoiling AGN, as well as the impact of GW recoil on BH/galaxy co-evolution.

We find that galaxy gas content and central escape speed at the time of the merger are the most important factors in determining the trajectory of a recoiling BH. However, the latter may vary rapidly during the time of BH coalescence, especially in gas-rich, nearly-equal-mass mergers. This adds an element of unpredictability to the recoiling BH dynamics.

Recoiling AGN may be distinguished from their stationary counterparts via kinematic or spatial offsets. In our models, recoiling AGN with $v_{\rm BH} > 500$~\kms~have~\tagn~up to a few tens of Myr; those with $R_{\rm BH} > 1$ kpc have~\tagn~up to $\sim 100$ Myr, though generally at lower luminosities. GW recoil also reduces the total lifetimes of bright, merger-triggered AGN, but may {\em increase} the low-luminosity~\tagn~if the BHs resume accretion upon settling back to the center. 

GW recoil may disrupt the coordinated growth of galaxies and BHs. Accordingly, the $M_{\rm BH}-\sigma_*$ relation resulting from our simulations with rapidly recoiling BHs has almost twice as much intrinsic scatter as that of an equivalent sample with stationary BHs. This additional scatter is an unavoidable outcome of GW recoil events. We have also shown that recoiling BHs may affect their host galaxies directly, owing to the displacement of AGN feedback from the galactic center. This extends the starburst phase and results in a merger remnant with a steeper central cusp.

In conclusion, our results indicate that GW recoil may be a non-negligible addition to the standard picture of BH/galaxy co-evolution. Further, the substantial lifetimes derived for some offset AGN suggest that observing such objects might be a real possibility in the near future.

\end{document}